\documentclass[final]{ws-p8-50x6-00}

\usepackage{graphics}
\usepackage{cite}

\def\ifmath#1{\relax\ifmmode #1\else $#1$\fi}%
\def\rb{\ifmath{{\mathrm{b}}}}
\def\rd{\ifmath{{\mathrm{d}}}}
\def\rA{\ifmath{{\mathrm{A}}}}
\def\rf{\ifmath{{\mathrm{f}}}}
\def\rF{\ifmath{{\mathrm{F}}}}
\def\rs{\ifmath{{\mathrm{s}}}}
\def\rt{\ifmath{{\mathrm{t}}}}
\def\rT{\ifmath{{\mathrm{T}}}}
\def\rW{\ifmath{{\mathrm{W}}}}

\def\pert{\ifmath{{\mathrm{pert}}}}
\def\uds{\ifmath{{\mathrm{uds}}}}
\def\QCD{\ifmath{{\mathrm{QCD}}}}
\def\PT{\ifmath{{\mathrm{PT}}}}

\newcommand{\epem}   {\ensuremath{\mathrm{e^+e^-}}}

\newcommand{\oaa}    {\ensuremath{\mathcal{O}(\alpha_s^2)}}
\newcommand{\oa}     {\ensuremath{\mathcal{O}(\alpha_s)}}
\newcommand{\bt}     {\ensuremath{B_\rT}}
\newcommand{\bw}     {\ensuremath{B_\rW}}

\newcommand{\thr}    {\ensuremath{1-T}}
\newcommand{\cp}     {\ensuremath{C}}
\newcommand{\cf}     {\ensuremath{C_\rF}}
\newcommand{\ca}     {\ensuremath{C_\rA}}
\newcommand{\tf}     {\ensuremath{T_\rF}}
\newcommand{\nf}     {\ensuremath{n_\rf}}
\newcommand{\anull}  {\ensuremath{\alpha_0}}
\newcommand{\bm}[1]  {\mbox{\boldmath\ensuremath{#1}}}
\newcommand{\momone}[1] {\mbox{\ensuremath{\langle#1\rangle}}}
\newcommand{\momtwo}[1] {\mbox{\ensuremath{\langle#1^2\rangle}}}
\newcommand{\as}     {\ensuremath{\alpha_\rs}}

\newcommand{\mui}    {\ensuremath{\mu_I}}

\newcommand{\roots}  {\ensuremath{\sqrt{s}}}
\newcommand{\mz}     {\ensuremath{M_{\mathrm{Z^0}}}}
\newcommand{\asmz}   {\ensuremath{\alpha_s(M_{\mathrm{Z^0}})}}
\newcommand{\ycut}   {\ensuremath{\mathrm{y_{cut}}}}
\newcommand{\znull}  {\ensuremath{\mathrm{Z^0}}}
\newcommand{\mb}     {\ensuremath{m_{\mathrm{b}}}}
\newcommand{\mbmz}   {\ensuremath{\mb(\mz)}}

\begin{document}

\title{ MULTIJET FINAL STATES IN \bm{\epem} ANNIHILATION }

\author{ S. Kluth }

\address{ Max-Planck-Institut f\"ur Physik, F\"ohringer Ring 6,
D-80805 Munich, Germany \\ 
E-mail: skluth@mppmu.mpg.de 
}

\maketitle

\abstracts{
We review the current status of analyses of multijet final states in
\epem\ annihilation. Results for jet observables from LEP~1, 
LEP~2 and from the reanalysis of the PETRA
experiment JADE will be presented.  A determination of the b-quark
mass using jet observables will be discussed and tests of power
correction models will be shown. Finally, determinations of
the QCD colour factors from an analysis of event shape distributions
at several energy points using power corrections will be discussed.  
}

\section{ Introduction }

Hadronic final states in \epem\ annihilation events are the subject of
many experimental and theoretical studies.  The structure of the
hadronic final states in \epem\ annihilation is characterised by the
presence of a small number of so-called jets, i.e. clearly separated
and collimated sprays of particles. As a consequence, hadronic events
may be classified by e.g. the number of jets after a jet finding
algorithm has been defined.  Alternative classification schemes are
event shape observables, where the reconstructed momenta of the final
state particles are combined in a way which characterises the
structure of the event in a single number.

Section~\ref{sec_jets} presents
results on jet production for centre-of-mass (cms) energies from 35
to 189~GeV, while section~\ref{sec_mb} describes a measurement of the
mass of the b-quark at a scale of \mz. Section~\ref{sec_pow} contains
a brief summary of experimental tests of power corrections.
Section~\ref{sec_col} shows a test of the gauge structure of QCD
and, finally, section~\ref{sec_summ} gives a summary of the report.

\section{ Jet Production from 35 to 189 GeV }
\label{sec_jets}

Jet production was studied comprehensively using data from the JADE
and OPAL experiments at the PETRA and LEP \epem\
colliders at cms energies from 35 to 189~GeV~\cite{OPALPR299}.
The large range of cms energies allows detailed studies of
scale dependent effects predicted by QCD. 

\begin{figure}[t]
\begin{tabular}{cc}
\includegraphics[width=0.5\textwidth]{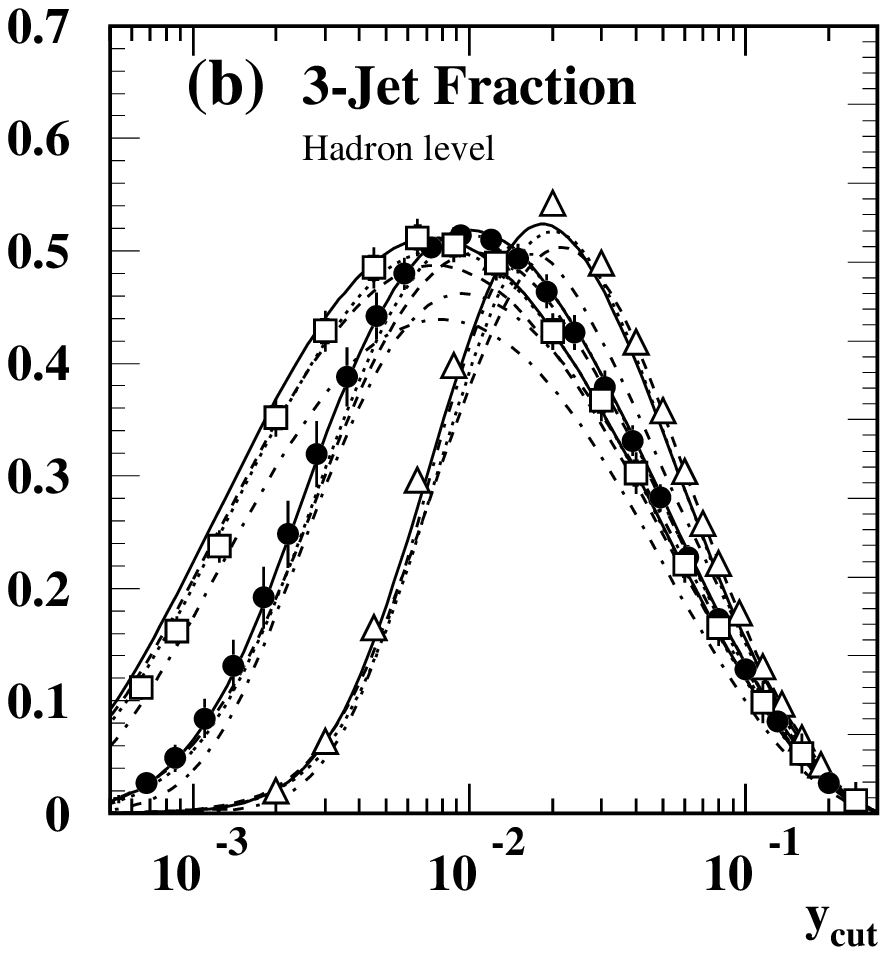} &
\includegraphics[width=0.5\textwidth]{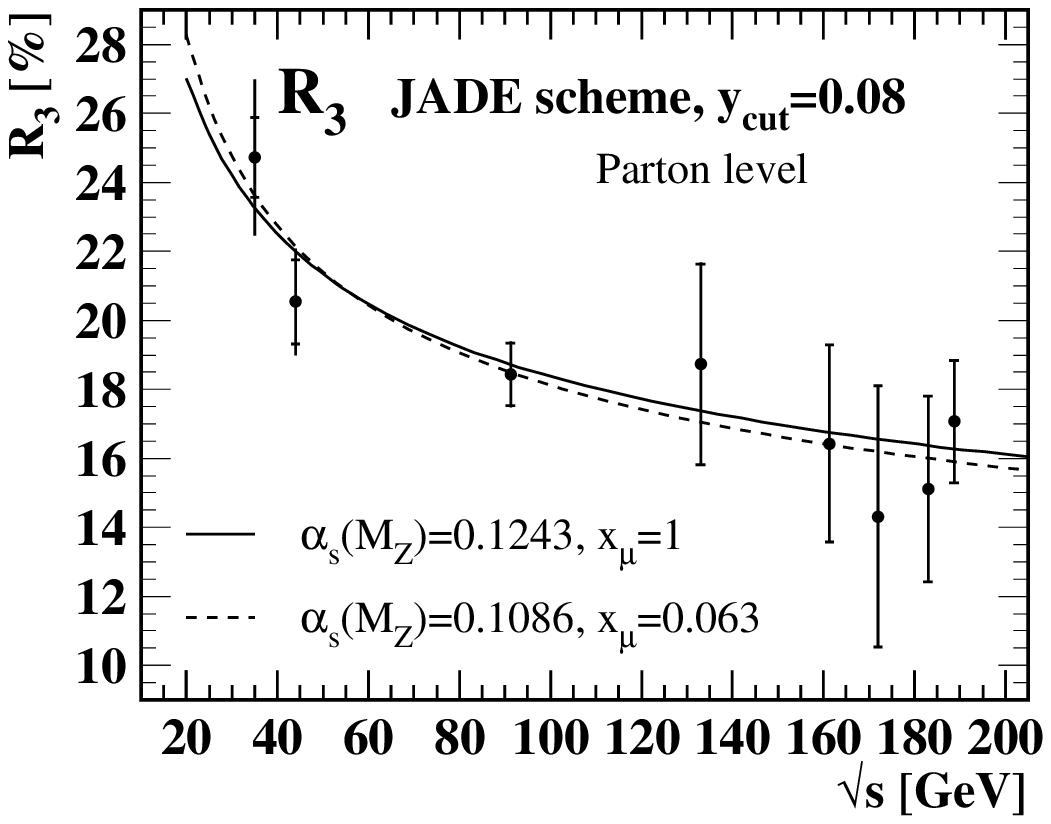} 
\end{tabular}
\caption[bla]{ The figure on the left shows the 3-jet fraction using
the JADE algorithm as a function of the jet resolution parameter
\ycut\ at $\roots=35$ (triangles), 91 (points) and 189~GeV (squares). 
The lines display Monte Carlo predictions by PYTHIA (solid), HERWIG
(dashed), ARIADNE (dotted) and COJETS (dash-dotted).  The figure on
the right displays the 3-jet fraction at $\ycut=0.08$ corrected for
experimental and hadronisation effects as a function of the cms energy
\roots\ \cite{OPALPR299}. }
\label{fig_jets}
\end{figure}

Figure~\ref{fig_jets} (left) shows the 3-jet fraction measured using
the JADE algorithm~\cite{jetsjade} at cms energies 35 (triangles) , 91
(points) and 189~GeV (squares) as functions of the jet resolution
parameter \ycut. In the JADE algorithm the invariant masses $m_{ij}$
of all pairs of final state particles are calculated and the pair with
the smallest value is combined into a pseudo-particle by adding their
4-vectors. This process is repeated until no invariant masses below a
cut value remain: $m_{ij}^2/s>\ycut$ for all $i,j$.  The decrease of
the 3-jet rate at large \ycut\ is clearly visible.  The distributions
are well described by Monte Carlo simulation programs tuned to OPAL
data recorded at $\roots=\mz$ \cite{OPALPR197,OPALPR141} including QCD
coherence effects in the parton shower, PYTHIA~\cite{jetset3} (solid),
HERWIG~\cite{herwig} (dashed) and ARIADNE~\cite{ariadne3} (dotted). A
simulation program without coherence effects,
COJETS~\cite{cojetstuning} (dash-dotted), describes the data less
well.

Figure~\ref{fig_jets} (right) shows the dependence of the 3-jet
fraction $R_3(\ycut=0.08)$ using the JADE algorithm as a function of
cms energy \roots, corrected for detector and for hadronisation
effects. In leading order, we have $R_3(\roots)\sim\as(\roots)$ for
$\ycut=0.08$~\cite{r3_prediction}; the data provide convincing
evidence for the running of \as\ as required by QCD.

\section{ Running b-Quark Mass } 
\label{sec_mb}

QCD predictions are generally calculated for massless quarks. This is
a good approximation for the light (${\cal O}(10-100)$~MeV) u-, d- and
s-quarks. However, for the heavy c- and b-quarks the masses are
comparable to energy scales where perturbative QCD calculations are
expected to be valid and thus quark mass effects may be
significant. Quark masses in the QCD Lagrangian are free
parameters like the strong coupling \as\ and have to be
renormalised to obtain finite predictions, see e.g. \cite{ellis96}.
The renormalised quark masses are expected to ``run'', i.e. to depend
on the energy scale of the process, because they must obey a
renormalisation group equation (RGE). The main effect of a heavy quark
mass in QCD is the suppression of gluon radiation from the heavy quark
which leads to the expectation of a reduced 3-jet rate.

\begin{figure}[t]
\begin{tabular}{cc}
\includegraphics[width=0.45\textwidth]{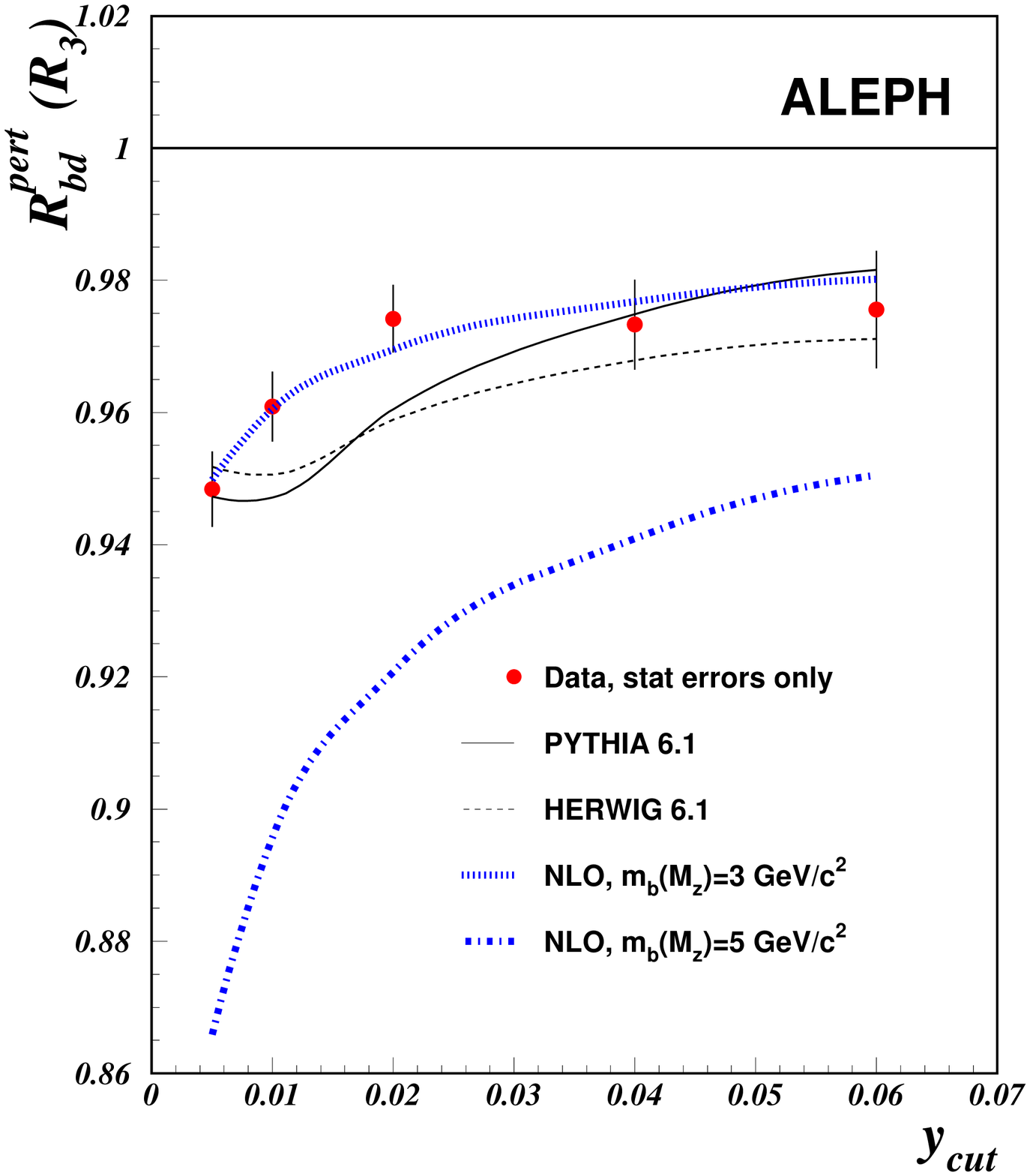} &
\includegraphics[width=0.45\textwidth]{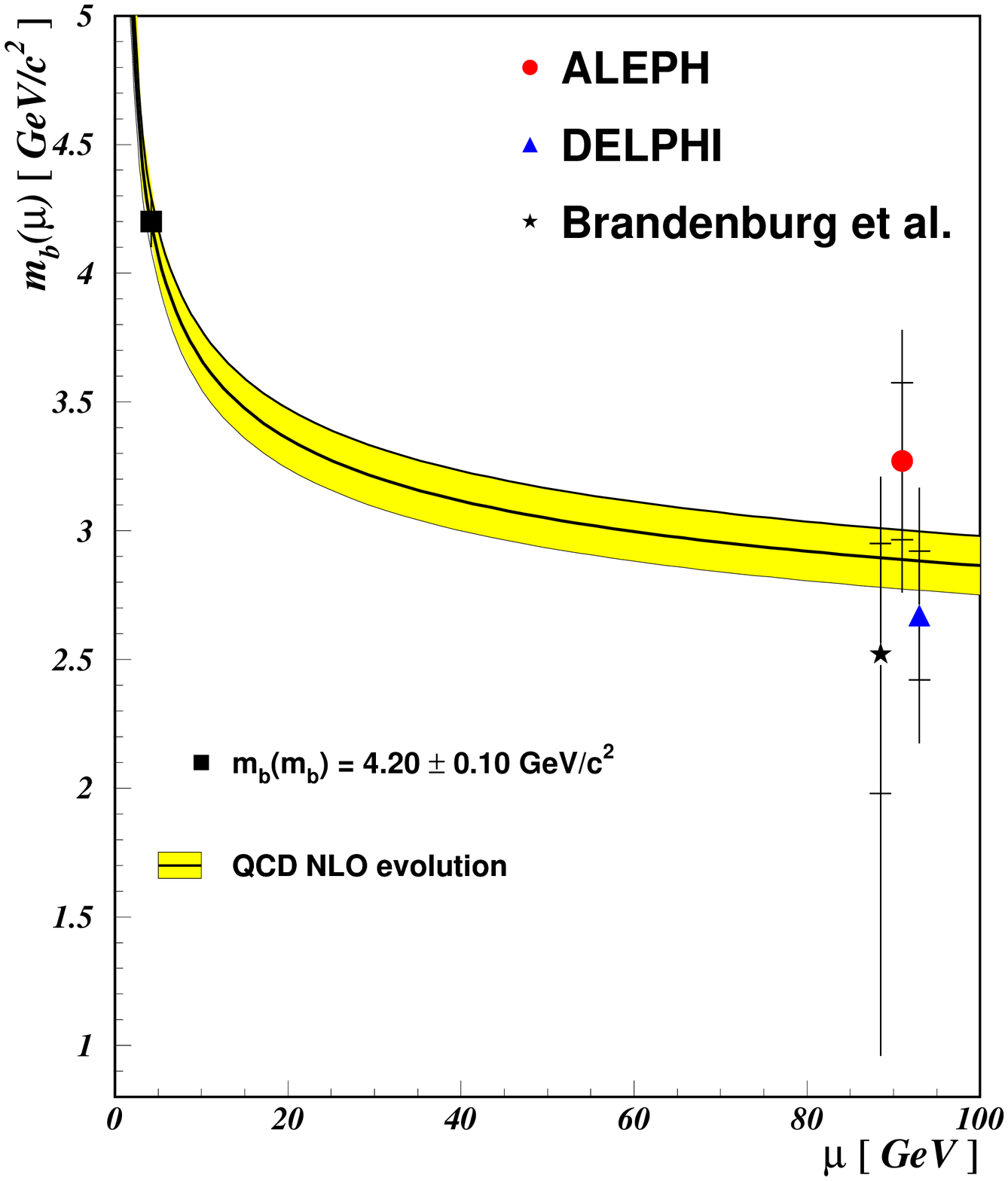} \\
\end{tabular}
\caption[bla]{ The figure on the left shows the ratio of 3-jet rates
in b- and d-quark events corrected to the parton level compared with
NLO QCD predictions and simulations. The figure on the right shows the
final result for \mbmz\ compared with other
measurements~\cite{Barate:2000ab}. } 
\label{fig_bmass}
\end{figure}

The ALEPH collaboration presented an analysis to test the prediction
of a running b-quark mass based on jet observables determined for b-
and light quark events~\cite{Barate:2000ab}. 
After correcting for experimental effects, b-tagging biases and
hadronisation effects, the ratio of the jet observable measurements in
b- over d-quark events is calculated, e.g. for the 3-jet rate $R_3$
one has $R^{\pert}_{\rb\rd}(R_3)=R_{3,\rb}/R_{3,\rd}$. Figure~\ref{fig_bmass}
(left) presents $R^{\pert}_{\rb\rd}(R_3)$ at several values of \ycut\
compared with NLO QCD predictions for $\mb=3$ and 5~GeV and Monte
Carlo simulations. The data clearly prefer the lower value of \mb\
while the simulations are in slight disagreement at low \ycut.

The final measurement of \mbmz\ is performed using the observable
with the smallest hadronisation corrections and systematic
uncertainties; this is the 1st moment of the differential 2-jet rate
distribution. The result $\mbmz=3.27\pm0.52$~GeV is presented in
figure~\ref{fig_bmass} (right) together with other measurements at the
\znull\ scale and at low scales. The QCD prediction of a running
b-quark mass starting from the value at low scale is in good agreement
with the measurements at \mz. Assuming the QCD description of heavy
quark effects to be correct the ratio 
$\as^\rb/\as^{\uds c}=0.997\pm0.009$
is determined and provides a precise test of the flavour independence
of the strong coupling $\as$.

\section{ Power Corrections }
\label{sec_pow}

Most measurements in QCD studies have to correct for the discrepancy
between perturbative QCD calculations and the quantities calculated
from the observed hadrons. These corrections are commonly carried out
using Monte Carlo models of the hadronisation process like
JETSET/PYTHIA, HERWIG or ARIADNE. An alternative approach to the
problem of hadronisation are analytical QCD based models of
hadronisation, the power corrections.

\begin{figure}[t]
\begin{tabular}{cc}
\includegraphics[width=0.4\textwidth]{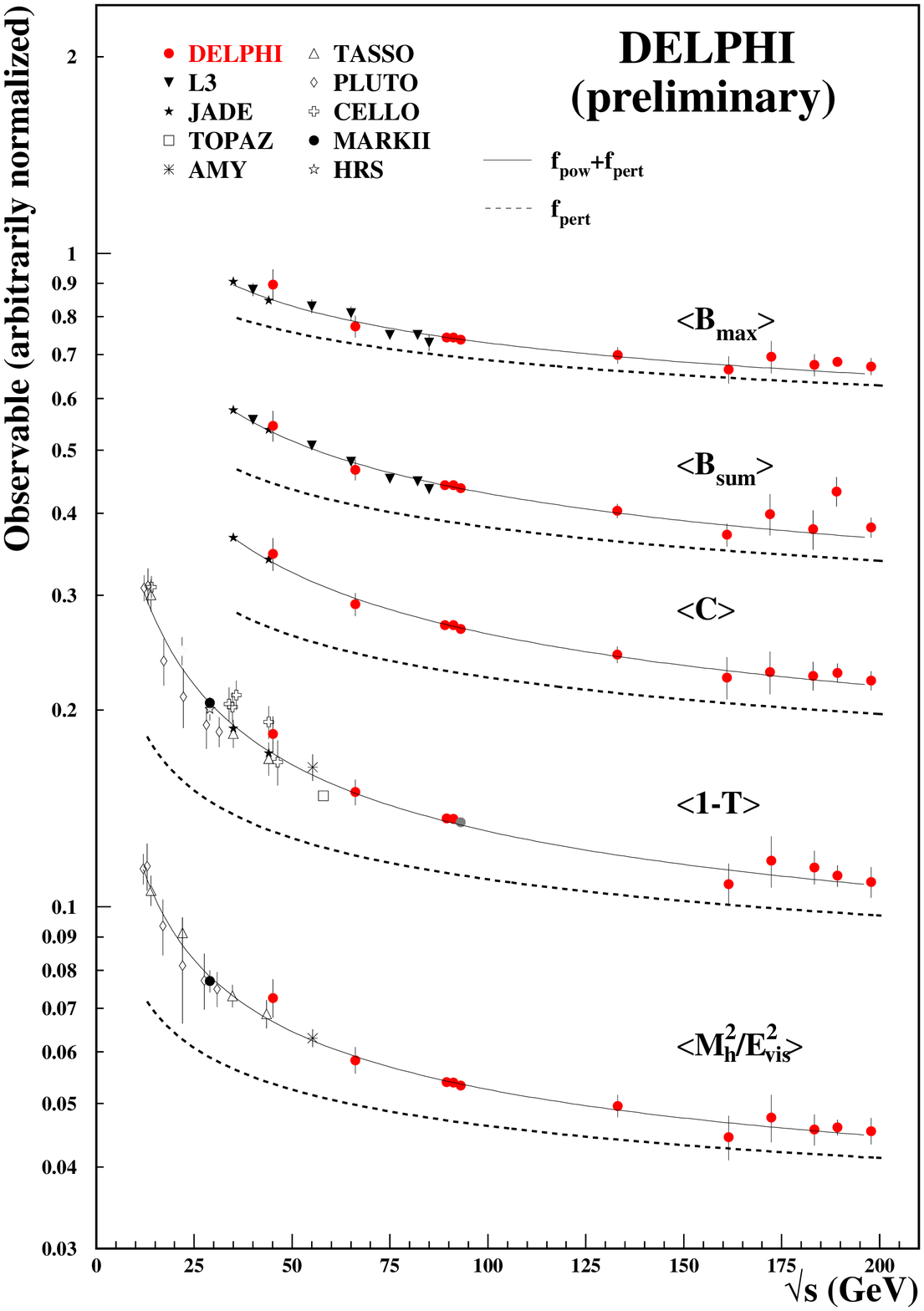} &
\includegraphics[width=0.45\textwidth]{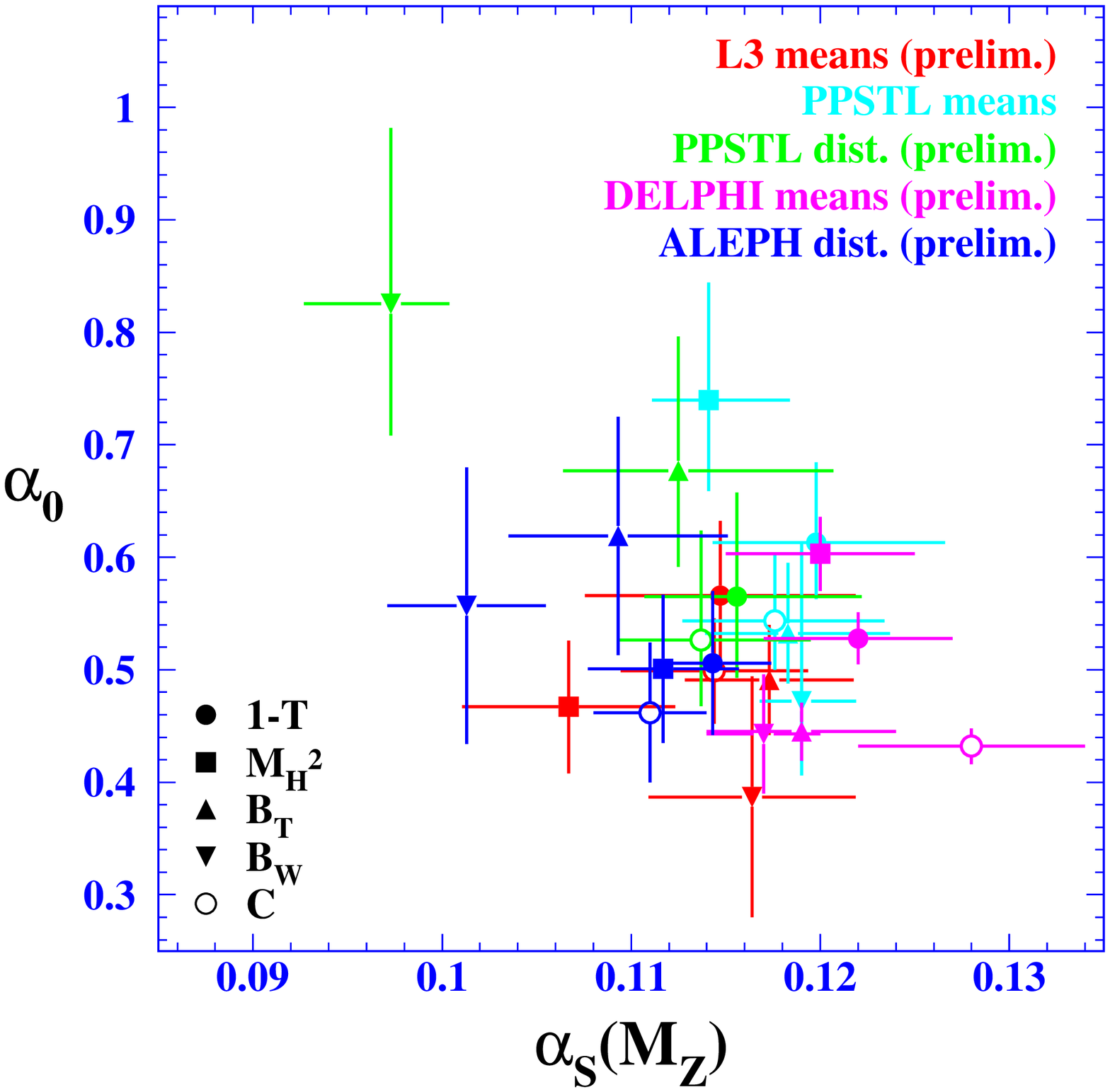} \\
\end{tabular}
\caption[bla]{ The figure on the left shows as solid lines \oaa\ QCD
fits with power corrections to 1st moments of event shape
observables~\cite{delphilep2data2}. The figure on the right shows 
a summary of all results for \anull\ and \asmz\ from various
analyses~\cite{l3lep2data2,jadec,jadedist,delphilep2data2,alephlep2data1}.
}
\label{fig_powc}
\end{figure}

In the Ansatz of Dokshitzer, Marchesini and Webber
(DMW)~\cite{dokshitzer95a}, the effects of gluons with transverse
momentum $k_\rt\sim\Lambda_{\QCD}$, so-called gluers, are calculated. The
model must assume that the strong coupling \as\ is finite in the
region of the Landau Pole leading to a new free parameter \anull\ in
the model: $\anull=\int_0^{\mui}\as(k)\rd k$. The variable \mui\ is the
infrared matching scale where non-perturbative and the perturbative
evolution of \as\ are merged.

For the differential distributions of the event shape observables
Thrust, Heavy Jet Mass, $C$-parameter and Total and Wide Jet Broadening,
the model predicts that hadronisation effects are described by a shift
of the perturbative prediction: $F(y)=F_{\PT}(y-c_yP)$ where $y$ is the
value of the observable~\cite{dokshitzer98b,dokshitzer99a}.  For the
1st and 2nd moment one obtains $\momone{y}=\momone{y}_{\PT}+c_yP$ and
$\momtwo{y}=\momtwo{y}_{\PT}+2\momone{y}_{\PT}c_yP+{\cal O}(1/Q^2)$. The
quantity $c_y$ depends on the observable while $P\sim
M\mui/Q\anull(\mui)$ is universal and the Milan factor $M$ takes
account of two-loop effects~\cite{dokshitzer98b}. The shift is
inversely proportional to the hard scale $Q$ usually identified with
the cms energy.

A study using 1st moments of event shape observables
by DELPHI is shown in figure~\ref{fig_powc} (left)
\cite{delphilep2data2} using DELPHI data from LEP 1 and 2 and data
from various experiments at lower energies. The fits of \oaa\ QCD
predictions with power corrections (solid lines) describe the data
well. The dashed lines represent the perturbative part, such that it
becomes apparent that hadronisation corrections are important even at
large cms energies. 

A direct test of the power corrections using differential
distributions of the event shape observables is presented
in~\cite{jadedist}. Data measured at $\roots=35$ to 183~GeV are fitted
simultaneously with only \asmz\ and \anull\ as free parameters. The
fitted predictions describe the data well within the fitted regions.

Results from power correction analyses for \asmz\ and \anull\ from
many recent analyses are summarised in figure~\ref{fig_powc} (right)
\cite{l3lep2data2,jadec,jadedist,delphilep2data2,alephlep2data1}.\footnote{
Results for \anull\ based on the old erroneous value of the Milan
factor $M=1.795$ have been scaled to correspond to the correct value
of $M=1.49$ \cite{dokshitzer99b}} The results for \asmz\ are
generally consistent with the world average value
$\asmz=0.119\pm0.003$~\cite{bethke00a} while the results for \anull\
are in agreement with each other at the 20\% level, as expected
theoretically~\cite{dokshitzer98b}. The results for \bw\
from distributions are not as consistent with universality of \anull\
as the other results.

\section{ QCD Colour Factors }
\label{sec_col}

A study of the QCD colour factors using fits of \oaa+NLLA QCD
predictions with power corrections to distributions of \thr, \cp, \bt\
and \bw\ was presented in~\cite{colrun}. The QCD colour factors
$\nf\tf$, \ca\ and \cf\ represent the relative contributions of the QCD
vertices of quark-pair production from a gluon, gluon-radiation of a
gluon (triple gluon vertex) and gluon radiation of a quark,
respectively. In the product $\nf\tf$, \nf\ is the number of active
quark flavours and \tf\ is the actual colour factor. 
The QCD colour factors are determined by the choice of
the gauge symmetry group, SU(3) in the case of QCD, and are expected
as $\nf=5$ for $\tf=1/2$, $\ca=3$ and $\cf=4/3$. 

The analysis follows~\cite{jadedist} but uses more recent data. The
dependence of the complete QCD predictions on the colour factors is
made explicit such that the colour factors \nf, \ca\ or \cf\ can be
varied in the fits, in addition to \asmz\ and \anull. The main
sensitivity comes from the running of \as, which in \oa\ reads
$\as(Q)=\as(\mu)/(1-2\beta_0\as(\mu)\ln(\mu/Q))$, where $\beta_0\sim
11\ca-2\nf$. Using the power correction calculations as the
hadronisation model reduces potential biases from hadronisation
corrections, because the power corrections depend explicitly on the
QCD colour factors.

\begin{figure}[t]
\begin{tabular}{cc}
\includegraphics[width=0.5\textwidth]{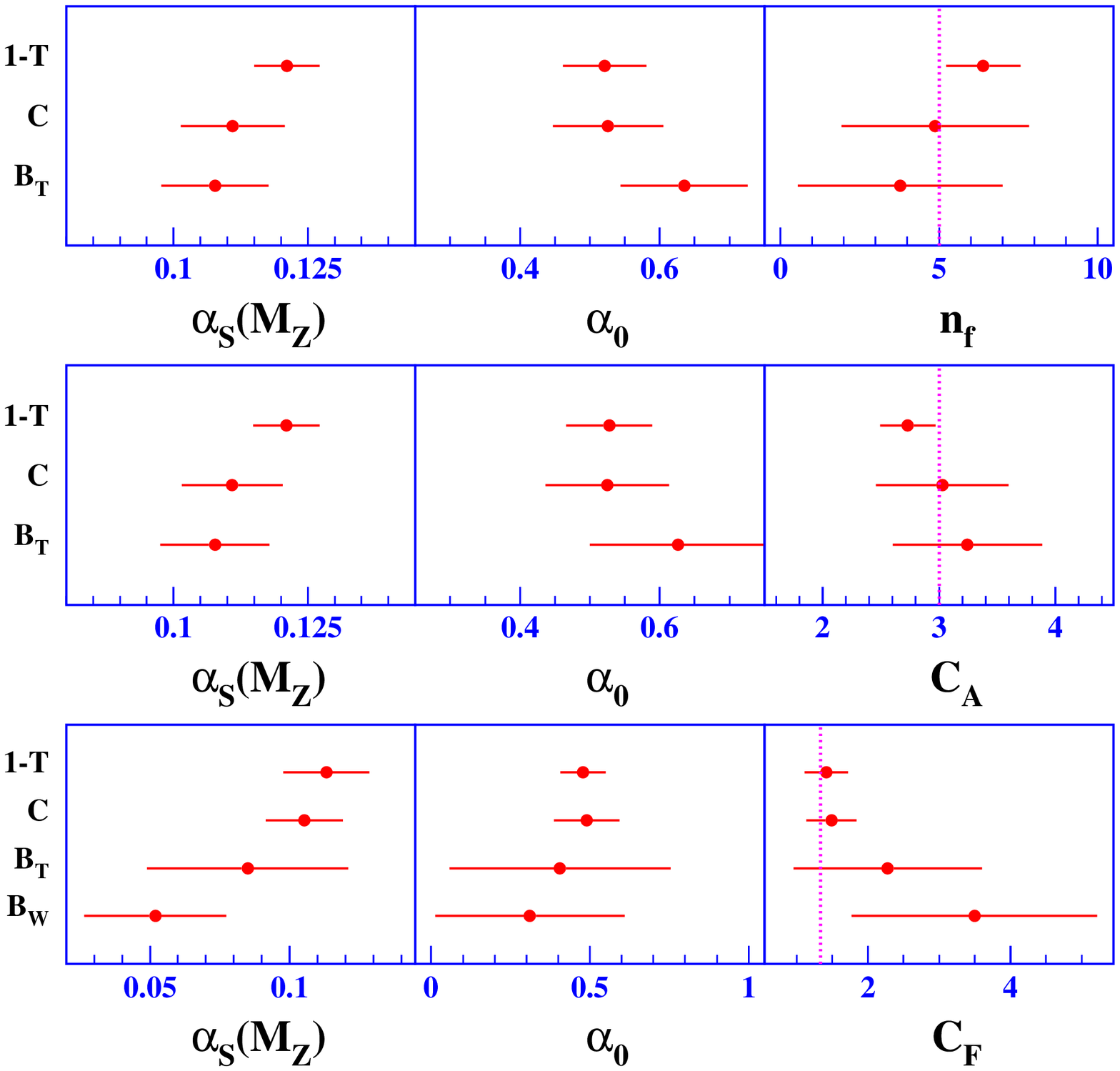} &
\includegraphics[width=0.5\textwidth]{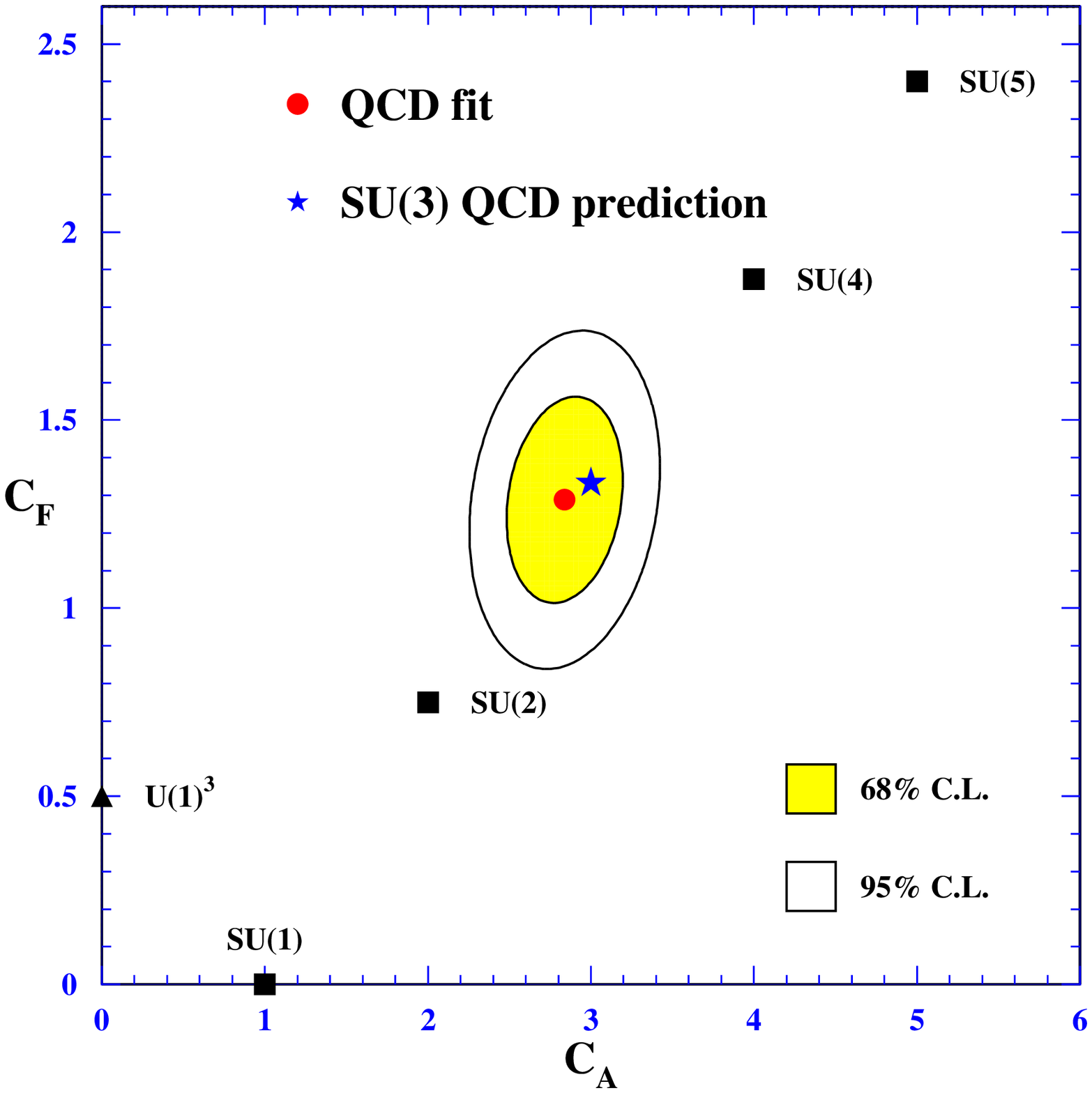} \\
\end{tabular}
\caption[bla]{ The figure on the left shows the results of fits to
event shape distributions with \asmz, \anull\ and one of the QCD
colour factors \nf, \ca\ or \cf\ as free parameters. The vertical
dotted lines indicate the expectations from standard QCD.  The figure
on the right shows the combined result of simultaneous fits of \asmz,
\anull, \ca\ and \cf\ to \thr\ and \cp\ \cite{colrun}. }
\label{fig_col}
\end{figure}

Figure~\ref{fig_col} (left) shows the results of fits with \asmz,
\anull\ and one of the colour factors \nf, \ca\ or \cf\ as free
parameters. The results for \asmz\ are consistent with the world
average, except the result from \bw\ when fitting \cf.  The results
for \anull\ are consistent with those presented in
section~\ref{sec_pow}. 
The results for the colour factors are
consistent with the SU(3) QCD expectation with five active flavours.
indicated by the vertical dotted lines.

Figure~\ref{fig_col} (right) displays the unweighted averages of
results of \thr\ and \cp\ from simultaneous fits with \asmz,
\ca\ and \cf\ as free parameters; the non-perturbative parameter
\anull\ was fixed at $\anull=0.543\pm0.058$. Also shown are the
expectations for various alternatives for the gauge symmetry group in
QCD, in particular $\mathrm{U}(1)^3$ is the representation for a
theory with three different neutral gauge bosons in direct analogy to
QED. The measurement agrees well with standard QCD with the SU(3)
symmetry group.

The analysis is complementary to the traditional approach of using
angular correlations in 4-jet final
states~\cite{aleph4jet2,4jetdelphi3,OPALPN430} and has similar total
uncertainties. Under the assumption that QCD based on the SU(3) gauge
symmetry group is the correct theory of strong interactions, the
analysis provides a successful consistency check of the power
correction model.

\section{ Summary }
\label{sec_summ}

We have shown experimental studies of jet production in \epem\
annihilation. Jet production as measured from PETRA to LEP~2 energies
is well described by QCD models and by perturbative QCD
calculations. A measurement of the b-quark mass at the \znull\ peak
provided evidence for the running of the b-quark mass as predicted by
QCD. Investigations of power corrections were discussed and it was
found that the model successfully predicts the hadronisation effects
for a number of event shape observables. The free non-perturbative parameter
\anull\ is observed to be universal within the theoretically expected
uncertainty of about 20\%. A measurement of the QCD colour factors
using power correction calculations was presented. This analysis is
complementary to traditional analyses of angular correlations in 4-jet
final states at the \znull\ peak and of similar accuracy.

The author would like to express his gratitude towards the organisers
of this meeting for a stimulating conference in a pleasant atmosphere.


\begin{thebibliography}{10}

\bibitem{OPALPR299}
JADE and OPAL Coll., G.~Abbiendi {\em et~al}, {\em Eur. Phys. J.} C {\bf
17}, 19 (2000). 

\bibitem{jetsjade}
JADE Coll., W.~Bartel {\em et~al}, {\em Z. Phys.} C {\bf 33}, 23 (1986).

\bibitem{OPALPR197}
OPAL Coll., K.~Ackerstaff {\em et~al}, {\em Z. Phys.} C {\bf 75}, 193 (1997).

\bibitem{OPALPR141}
OPAL Coll., G.~Alexander {\em et~al}, {\em Z. Phys.} C {\bf 69}, 543 (1996).

\bibitem{jetset3}
T.~Sj{\"o}strand, {\em Comput. Phys. Commun.} {\bf 82}, 74 (1994).

\bibitem{ariadne3}
L.~L{\"o}nnblad, {\em Comput. Phys. Commun.} {\bf 71}, 15 (1992). 

\bibitem{herwig}
G.~Marchesini {\em et~al}, {\em Comput. Phys. Commun.} {\bf 67}, 465 (1992).

\bibitem{cojetstuning}
P.~Mazzanti and R.~Odorico, {\em Nucl. Phys.} B {\bf 394}, 267 (1993).

\bibitem{r3_prediction}
S.~Bethke, Z.~Kunszt, D.E. Soper and W.J. Stirling, {\em Nucl. Phys.} 
B {\bf 370}, 310 (1992).

\bibitem{durham}
S.~Catani {\em et~al}, {\em Phys. Lett.} B {\bf 269}, 432 (1991).

\bibitem{dokshitzer97b}
Yu.~L. Dokshitzer, G.~D. Leder, S.~Moretti and B.~R. Webber, {\em JHEP} 
{\bf 8}, 1 (1997).

\bibitem{bethke00a}
S.~Bethke, {\em J. Phys.} G {\bf 26}, R27 (2000).

\bibitem{ellis96}
R.K. Ellis, W.J. Stirling and B.R. Webber in {\em QCD and Collider Physics}. 
Vol.~8
  of Cambridge Monographs on Particle Physics, Nuclear Physics and Cosmology,
  Cambridge University Press (1996)

\bibitem{Barate:2000ab}
ALEPH Coll., R.~Barate {\em et~al}, CERN-EP/2000-093 (2000)

\bibitem{delphilep2data2}
DELPHI Coll., P.~Abreu {\em et al}, 
DELPHI 2000-022 CONF 343 (2000), Submitted to
  XXXVth Rencontres de Moriond, March 18 - March 25, Les Arcs, France

\bibitem{l3lep2data2}
L3 Coll., M.~Acciarri {\em et~al}, L3 note 2555 (2000), Submitted to XXXth
  International Conference on High Energy Physics, July 27 - August 2, Osaka,
  Japan

\bibitem{jadec}
JADE Coll., O.~Biebel, P.~A.~Movilla Fernandez, S.~Bethke {\em et~al}, 
{\em Phys. Lett.} B {\bf 459}, 326 (1999).

\bibitem{jadedist}
P.~A.~Movilla Fernandez, O.~Biebel and S.~Bethke, {\it PITHA} 99/21 (1999).

\bibitem{alephlep2data1}
ALEPH Coll., D.~Abbaneo {\em et~al}, ALEPH 2000-044 (2000), Submitted to XXXth
  International Conference on High Energy Physics, July 27 - August 2, Osaka,
  Japan

\bibitem{dokshitzer95a}
Yu.L. Dokshitzer, G.~Marchesini and B.R. Webber, {\em Nucl. Phys.} 
B {\bf 469}, 93 (1996).

\bibitem{dokshitzer98b}
Yu.~L. Dokshitzer, A.~Lucenti, G.~Marchesini and G.~P. Salam, 
{\em JHEP} {\bf 5}, 3 (1998).

\bibitem{dokshitzer99a}
Yu.~L. Dokshitzer, G.~Marchesini and G.~P. Salam, 
{\em Eur. Phys. J.} direct C {\bf 3}, 1 (1999).

\bibitem{dokshitzer99b}
Yu.~L. Dokshitzer: hep-ph/9911299 (1999), Invited talk at 11th Rencontres de
  Blois: Frontiers of Matter, Chateau de Blois, France, 28 Jun - 3 Jul 1999

\bibitem{colrun}
S.~Kluth {\em et~al}, MPI-PhE/2000-19 (2000).

\bibitem{aleph4jet2}
ALEPH Coll., R.~Barate {\em et~al}, {\em Z. Phys.} C {\bf 76}, 1 (1997).

\bibitem{4jetdelphi3}
DELPHI Coll., P.~Abreu {\em et~al}, {\em Phys. Lett.} B {\bf 414}, 401 (1997).

\bibitem{OPALPN430}
OPAL Coll., G.~Abbiendi {\em et~al}, OPAL physics note PN430 (2000), 
unpublished.

\end{thebibliography}
\end{document}